\documentclass[12pt,preprint]{aastex}

\def\masyr{{\rm mas}\,{\rm yr}^{-1}}

\def\nltt{{\rm NLTT}}

\def\lim{{\rm lim}}

\begin{document}

\title{Completeness of USNO-B for High Proper-Motion Stars}

\author{Andrew Gould}
\affil{Department of Astronomy, The Ohio State University,
140 W.\ 18th Ave., Columbus, OH 43210}
\authoremail
{gould@astronomy.ohio-state.edu}

\singlespace

\begin{abstract}

I test the completeness of USNO-B detections of high proper-motion 
($\mu>180\,\masyr$) stars and the accuracy of its measurements by
comparing them to the revised NLTT (rNLTT) catalog of Salim \& Gould.
For $14.5<V<18.5$, only 6\% of such stars are missing from USNO-B 
while another 4\% have large errors, mostly too large to be useful.
Including both classes, incompleteness is 10\%.
These fractions rise toward both brighter and fainter magnitudes.
Incompleteness rises with proper motion to $\sim 30\%$ at 
$\mu=1''\,\rm yr^{-1}$.  It also rises to $\sim 35\%$ at the Galactic
plane, although this is only determined for relatively bright stars
$V\la 14$.  For binaries, incompleteness rises from 10\% at separations
of $30''$ to 47\% at $10''$.  The proper-motion errors reported internally
by USNO-B are generally correct.  However, there is floor of
$\sigma_\mu\sim 4\,\masyr$ below which the reported errors should not
be taken at face value.  The small number of stars with relatively large 
reported errors ($\sigma_\mu \ga 20\,\masyr$) may actually have still
larger errors than tabulated.

\end{abstract}
\keywords{astrometry --  methods: statistical}
 
\section{Introduction
\label{sec:intro}}

Until recently, the {\it New Luyten Two-Tenths} (NLTT) catalog
\citep{luy} was the only all-sky catalog of high proper-motion
stars extending to faint magnitudes.  Since its completion more
than two decades ago, after a lifetime of work by Luyten, the
NLTT has served as an invaluable source of candidate subdwarfs,
white dwarfs, and nearby stars, as well as the basis for statistical
studies of Galactic populations.  

However, the NLTT does suffer from several significant shortcomings.
First, while most NLTT positions are accurate to the
nominal catalog precision of ($1\,{\rm s}\times 6''$), 
there is a tail of errors
extending out to several arcmin and beyond \citep{bright,faint}.  This
means that a significant fraction of NLTT stars cannot easily be located.
Second, the photographic magnitudes (and especially colors) given by NLTT
are not precise enough to reliably classify stars using a reduced
proper motion (RPM) diagram \citep{rpm}.  Third, NLTT is seriously
incomplete at faint magnitudes both close to the Galactic plane and in the 
areas south of the first Palomar Observatory Sky Survey ($\delta<-33^\circ$), 
especially $\delta <-45^\circ$.  Moreover, its completeness
over the rest of the sky is not definitively known.  

Since these shortcomings significantly limit NLTT's application to various 
problems, substantial efforts have been made to rectify them.  For
example,  \citet{reidcruz} cross identified NLTT stars with
2MASS \citep{2mass} entries, while \citet{bakos} directly identified
stars from the {\it Luyten Half-Second} (LHS) catalog \citep{lhs} on
the Digitized Sky Survey (DSS).  LHS is a subset of NLTT containing 
essentially all NLTT stars with proper motions $\mu>500\,\masyr$ as
well as some slightly slower than this limit.  The nominal threshold for
NLTT is $\mu\geq
180\,\masyr$.  \citet{bright} cross identified NLTT bright stars
with three modern position and proper motion (PPM) catalogs,
Hipparcos \citep{hip}, Tycho-2 \citep{t2}, and Starnet \citep{starnet},
while \citet{faint} cross identified NLTT faint stars with counterparts
from USNO-A \citep{usnoa1,usnoa2} and 2MASS.  Together these comprise
the revised NLTT catalog (rNLTT).  To the extent that one can
find 2MASS counterparts of NLTT stars, they can be placed on an 
optical/infrared color-magnitude diagram (CMD), which allows much more
secure RPM classification \citep{rpm}.  

Finally, several groups have worked to estimate the completeness of NLTT
or, in some cases, improve it.  \citet{nreid} made an independent
search for stars exceeding NLTT's proper-motion threshold over
a single high-latitude Schmidt plate.  \citet{monet} searched for high
proper-motion ($\mu>400\,\masyr$) stars to faint magnitudes toward 
1378 deg$^2$.  \citet{bright} directly measured the completeness of bright
($V\la 11$) stars by cross-identifying Tycho-2 and Hipparcos stars.
\citet{flynn} estimated NLTT completeness based on its
internal properties, while I  made a parameterized measurement
of NLTT completeness as a function of magnitude as a byproduct of my 
measurement of the properties of halo stars \citep{parms}.  In the appendix to
that paper, I argued that all the available evidence is consistent
with the following picture.  NLTT is virtually 100\% complete for $V\la 11$
except near the plane where it falls to 75\%.  At fainter magnitudes
(and away from the plane),
completeness falls gradually to $\sim 50\%$ at $V\sim 18$.  Even at these
faint magnitudes, completeness is much higher for the small subset of
high proper-motion stars, $\mu>500\,\masyr$.  To rectify the 
relatively high incompleteness levels toward the plane, \citet{lepine}
have undertaken a new intensive proper-motion search in this region.

With the publication of the 
long-awaited USNO-B1.0 \citep{usnob}, the prospects for
improving, or perhaps superseding the NLTT have taken a sharp turn for the
better.  USNO-B1.0 is an all-sky PPM catalog based on deep photographic plates
taken over about four decades in the north $(\delta>-33^\circ)$ and
two decades in the south.  It contains $10^9$ entries, which not only
specify the adopted PPM (and errors), but also the astrometric and photometric 
measurements at all the epochs that were combined to make these determinations.

However, although USNO-B1.0 is a monumental undertaking, it cannot be used 
immediately to assemble a catalog of high proper-motion stars.  By deliberate
construction, USNO-B1.0 contains entries for {\it all} sets of detections
within $30''$ that can reasonably be matched to a straight-line motion over 
time (and which have not previously been matched at higher confidence to
other such linear ensembles of detections).  
These identifications were effected without making use of any additional 
information, such as photometry. This means that USNO-B1.0
inevitably contains a large number of false high proper-motion entries.
At high latitude, it contains about 200 times more entries than NLTT,
meaning that $\ga 99\%$ of these entries are spurious, formed of incorrect
matching of unassociated stars or even of plate artifacts.  As \citet{usnob}
emphasize, this contamination is deliberate.  The catalog's compilers 
sought to allow its users to sift through all possible associations to
find the most possible genuine high proper-motion stars.  I will discuss
several ideas for how to go about doing this in \S~\ref{sec:discuss}.
However, before any of these are actually implemented 
(or even precisely defined) it is essential to understand the properties
of the catalog itself, most importantly its completeness.

In this paper, I study the completeness of USNO-B by comparing it to the
revised NLTT (rNLTT) as compiled by \citet{bright} and \citet{faint}.
I perform two complementary comparisons, one to rNLTT binaries and one
to single stars.  Binaries in rNLTT are extremely well understood because
the supplementary notes to NLTT give information on the separation 
and the position angle of the components.  These permit an extra check on the
reality of the rNLTT identifications.  Binaries also constitute a
subset of NLTT that is of interest in its own
right.  Hence, if NLTT is eventually extended using USNO-B, it will be
important to understand USNO-B's completeness as a function of binary
separation.  I will show that this completeness is indeed a strong
function of separation.  Given this fact, binaries may seem a poor proxy
for the single stars that constitute the vast majority of NLTT entries.
I handle this problem in two ways.  First, I examine the binaries themselves
at very wide separation where USNO-B is unlikely to be affected by 
confusion caused by companions.  Second, I examine single stars from rNLTT
directly.  Because they are much more numerous, single stars also permit
study of completeness as functions of apparent magnitude, proper motion, 
and Galactic latitude.  Single stars
do have the drawback that their identifications in rNLTT are not quite as
secure as are those of binaries, and this fact ultimately circumscribes
the conclusions that can be drawn from studying them.  Nevertheless, the
implications of the two comparisons are broadly consistent, and thus each
lends credence to the other.

In \S~\ref{sec:rnltt}, I briefly review the properties of the rNLTT.
In \S~\ref{sec:bin} and \S~\ref{sec:sing}, I analyze the 
completeness of USNO-B relative to, respectively, binary and single-star
samples from rNLTT.  In \S~\ref{sec:errors}, I determine the accuracies
of the USNO-B proper-motion measurements and compare these to its
internal error estimates.  Finally, in \S~\ref{sec:discuss}, I discuss the
prospects for combining USNO-B with other catalogs to obtain a clean,
relatively complete sample of high proper-motion stars.

\section{Review of the Revised NLTT (rNLTT)
\label{sec:rnltt}}

The construction of the rNLTT was a huge undertaking, which is laboriously
documented in \citet{bright} and \cite{faint}.  Here I review the basic
features of this catalog as they affect the current paper.

As mentioned in \S~\ref{sec:intro}, bright $(V\la 11)$ NLTT stars are matched 
to bright PPM catalogs and faint NLTT stars are (mostly) matched to
pairs of stars from USNO-A and the 2MASS
 Second Incremental Data Release.  Bright stars will not be a major
concern here because  USNO-B does not independently locate bright stars.
Rather, it simply incorporates Tycho-2 entries directly into the catalog.
Hence, rNLTT and USNO-B typically have the same bright-star entries
since they both come from the same source.  Bright stars are of interest
only when they are components of common proper motion (CPM) binaries.
In this case, the bright star is used solely to help establish the
reality of the identification of its faint companion.

While the great majority of faint stars are identified as USNO-A/2MASS
pairs, some are found through several other channels.  First, if
no plausible pair is found within $2'$ of the position given by NLTT,
but a 2MASS star is found within $12''$ of the position predicted by
NLTT, then it is accepted as a match (provided it does not have a 
USNO-A counterpart
at the same position, which would indicate that it is not a high proper-motion
star).  The theory is that the NLTT star is missing from USNO-A, which can
happen due to crowding, faintness, 
non-detection on blue plates, or other reasons that are not discernible.
Second, if there is a unique USNO-A star within a $16''\times 8''$ 
rectangle of the NLTT position but no corresponding 2MASS star, then it
also is accepted.  In this case it is assumed that the star is missing
from 2MASS.  Third, if a star is located either in a PPM catalog or as
a USNO-A/2MASS pair, but its binary companion is not, then the 2MASS
catalog is searched at the relative offset predicted by the NLTT notes
on binaries.  Companions within a few arcsec of the predicted position
are accepted, provided they do not have USNO-A counterparts (which again
would indicate that they are not high proper-motion stars).  Finally,
a small number of stars are identified by other means, for example
by confirming identifications of \cite{bakos} using 2MASS.

At this point rNLTT does not cover the whole sky.  For faint stars,
it is restricted to the 44\% of the sky covered by intersection of the 
the first Palomar Observatory Sky Survey (POSS I), roughly $\delta>-33^\circ$,
and the 2MASS Second Incremental Release.  From the standpoint of doing
statistical studies, this is not a significant limitation 
(e.g., \citealt{parms}), but it does need to be kept in mind.

The rNLTT is about 97\% complete relative to NLTT up to $R_\nltt\sim 17$
(roughly $V\sim 18$), and is 95\% complete one magnitude fainter.  However,
in this last magnitude, the completeness is heavily dependent on
single-catalog detections, which are less secure, while at brighter
mags the overwhelming majority of entries are detected in both USNO-A
and 2MASS (see figure 13 of \citealt{faint}).  This is important because,
while the overall false identification rate in rNLTT is only 1\%,
these errors are probably concentrated in the single-catalog identifications.

The proper-motion errors in rNLTT are about $5.5\,\masyr$, but this excludes
the $3\,\sigma$ outliers (about 5\%).  However, in the present context
it is important to note that very few of these outliers lie beyond
$30\,\masyr$ (see figure 10 of \citealt{faint}).  While these errors
are based on relatively bright ($V\la 12$) USNO-A/2MASS detections, a
separate test on CPM binaries (but not reported by \citealt{faint})
shows that the errors do not evolve strongly with magnitude.

The epoch 2000 positions (and hence the position errors) come basically
from 2MASS.  These are typically 130 mas, but do deteriorate for the
faintest stars $V\sim 19$ to about 300 mas or so.  This will be important
when trying to evaluate the completeness of USNO-B in this faintest bin.

Finally, I review the characteristics of rNLTT binaries.  
The (narrow-angle) proper-motion
errors for these are only $3\,\masyr$.  The vector separations given by
the NLTT notes are generally accurate to $<1''$ (see figure 8 of 
\citealt{faint}).
This permits an additional check on all stars that are members of binaries.
When the two components are found separately they must have the offset
predicted by the NLTT notes.  There are few enough candidate identifications
that do not that satisfy this condition that these
can be individually checked to insure that they are
correct.  For companions that are not identified independently, the
precise prediction for each one's 
position allows identification with good confidence
(and additional checks for the handful of suspicious cases).  Thus,
the 2MASS-only binary detections can be accepted with much higher confidence 
than is the case for single stars.

\section{USNO-B Completeness from Binaries
\label{sec:bin}}

To determine USNO-B completeness, I consider a restricted set of rNLTT
binaries.  First, I examine only binaries for which both components are
detected and for which these components 
are separated by at least $\Delta\theta\geq 10''$.
Closer binaries can be blended in USNO-A, which leads to unreliable
proper-motion measurements.  Second, I exclude components of wider binaries
that are themselves members of closer binaries.  This includes both
triple systems recognized by NLTT as well as NLTT ``single stars'' that
are resolved by the Tycho Double Star Catalog (TDSC, \citealt{tdsc}).
Fourth, I exclude all binaries for which both components 
have PPM identifications.  As mentioned above, USNO-B generally takes over the
Tycho-2 proper motion for these stars, so they contain no information
on USNO-B completeness.  Next, I exclude all binaries for which one component
has a USNO-A-only identification because the additional checks discussed at the
end of \S~\ref{sec:rnltt} are not available for these stars.  This leaves
549 pairs, of which 85 have at least one component that is detected
in 2MASS only.  I put these latter aside for the moment and return to them
later.  Hence, 464 pairs remain.  

To match these stars, I first search within $1''$ of their predicted
position in USNO-B.  There are only 12 multiple matches and these
are easily resolved by hand.  For the nonmatches, I increase the
search radius to $3''$, which leads to 13 additional matches, all
but one within $2''$.  Of the 464 pairs, 378 have USNO-B identifications
for both components.

Figure \ref{fig:pmdif} shows the difference in the proper motion of the
two components of the 378 binaries as measured by USNO-B and rNLTT.  
Of course to make this figure, I have included only binaries with
independent detections of both components in each catalog.  
Figure \ref{fig:pmdif} has three notable features: a diagonal plume,
a vertical plume, and a dense knot close to the origin.  The diagonal
plume consists of pairs that are not physical pairs and were misidentified
as such by Luyten.  Both rNLTT and USNO-B agree on this verdict.  The
vertical plume consists of pairs that Luyten correctly identified as physical
(or at any rate correctly identified as having very similar proper motions)
as confirmed by rNLTT.  The fact that USNO-B finds large proper-motion
differences between the components indicates that one or both measurements
are erroneous.  Note that there is no similar horizontal plume, which 
would consist of real CPM binaries that were confirmed as such by USNO-B
but mismeasured by rNLTT.  Finally, the dense knot consists of binaries
that are confirmed by USNO-B and rNLTT to be physical.  The inset shows that,
while some of this relative motion may be real orbital motion, 
most of the scatter is
uncorrelated between USNO-B and rNLTT and so is probably due to measurement
error.  The bold curve shows the contour to the left of which I 
conclude that
one of the USNO-B measurements is in error.  After inspection of these,
I find that in one case, both are in error.  There are three
points that do not lie within this contour, nor in the dense knot, nor
in the diagonal plume.  In the following, I will assume that these are
accurate measurements.

Figure \ref{fig:sep1} shows the individual components of the 464
binaries analyzed above (i.e., all pairs for which both components
have independent rNLTT proper-motion measurements but excluding pairs
for which both components were detected in PPM catalogs).  These
464 pairs contain a total of 158 PPM stars, so the figure shows
$2\times 464 - 158 = 770$ stars.  Of these, 33 (shown by solid circles
with error bars) are erroneously measured components from the pairs lying 
inside the contour in
Figure \ref{fig:pmdif}.  The error bars indicate the proper-motion difference
(i.e., the ordinate from Fig.~\ref{fig:pmdif})
in units of $200\,\masyr$ from USNO-B measurements.  The solid circles
without error bars represent the 96 stars that were not detected in USNO-B.
When a star was detected in USNO-B but its companion was one of the 96 that
were not, I checked
its proper motion against rNLTT.  The seven cases for which the difference
exceeded $25\,\masyr$ are shown by a star.
The remaining 639 stars,
which are deemed good proper-motion measurements, are shown by small dots.
The $V$ magnitudes and $\Delta\theta$ separations are taken from rNLTT.

Figure \ref{fig:sep1} has two striking features.  First, the ``problem stars''
(poor measurements and non-matches) are concentrated at separations
$\Delta\theta<60''$.  Second, in the region $\Delta\theta>60''$, they
are concentrated at bright magnitudes $V<13.5$.  Hence, one may already
conjecture that these problems are due to three effects, crowding (at
very close separations), confusion (at moderate separations) and 
saturation at bright magnitudes.

To obtain a more complete picture, it is necessary to include the
2MASS-only detections from rNLTT.  There are 85 such binaries of which
39 have one component identified from PPM catalogs.  Hence, there are an
additional $2\times 85 -39=131$ stars for which USNO-B completeness and
accuracy can in principle be tested.  Of these, 59 are missing from USNO-B.
When the remaining matches agree between USNO-B and rNLTT, I conclude that
the USNO-B measurement is correct.  There is a serious disagreement in only one
case.  By comparing the separations as listed in the NLTT notes and
rNLTT, I find that the USNO-B proper motion is incorrect.  In all the
cases for which components are missing from USNO-B, comparison of the NLTT
and rNLTT separations confirms that the
proper motions of the two stars is similar enough that the 2MASS-only
companion should have been recovered in USNO-B (if it had been there).

Figure \ref{fig:sep2} shows the components of these rNLTT binaries
that were (solid circles) and were not (points) detected by USNO-B.
The one USNO-B detection with a bad proper motion is indicated with
an error bar.  Both the underlying
rNLTT sample and the USNO-B non-detections are heavily concentrated
at small separations and faint magnitudes.  This is reasonable.
Most stars that are missing from USNO-A are either too crowded
or too faint to be reliably detected.  Proximity to a companion
is one possible cause of crowding.   These same features would make
them difficult to detect in USNO-B (which is working off of the same
and similar plate material).  However, occasionally stars are missing
from USNO-A for no discernible reason.  If these stars were also preferentially
missing from USNO-B, then one would expect a high relative density of 
solid circles among the relatively faint $(13.5<V<19)$, well-separated
($\Delta\theta>60''$) stars, which appear so reliably detected in Figure
\ref{fig:sep1}.  That this is not the case will be important in the 
interpretation of the single-star results reported in \S~\ref{sec:sing}.

Figure \ref{fig:cumsep} shows the cumulative distribution of ``problem
stars'' (relative to all stars) as a function of binary separation.  Problem
stars are defined as those that were not detected in USNO-B, or whose proper
motions are significantly in error (solid circles and stars in Figs.\
\ref{fig:sep1} and \ref{fig:sep2}).  This cumulative distribution 
is restricted to stars $V>13.5$.
For $\Delta\theta\geq 60''$, the problem rate is $10/208=5\%$, while for
$30''\leq\Delta\theta<60''$, the rate is $17/172=10\%$ and for
$10''\leq\Delta\theta<30''$, it is $97/361=27\%$.  Apparently, the detection 
problems increase substantially for $\Delta\theta<60''$, and do so much
more dramatically for $\Delta\theta<30''$.

These results are somewhat puzzling.  While problems at 
$\Delta\theta=30''$ are easy to understand, it is much more difficult
to account for problems at $\Delta\theta=60''$.  USNO-B attempts to
associate unmatched stars at displacements up to $30''$ of their counterpart's
position at a previous epoch.  Hence,
a binary companion at $30''$ (which would of course be initially unmatched
by the preliminary algorithm that matches nonmoving stars) could easily
cause confusion.  Stars at somewhat greater separation could still create
problems if their proper motions moved them within $30''$.  However, the
problem stars have rather typical proper motions for NLTT, with almost
all having $\mu<500\,\masyr$.  Such proper motions would carry them only
$15''$ (and generally much less) over typical 30-year epoch differences.
One is therefore led to suspect some sort of selection effect in the
procedure I have adopted.  However, I am unable to think of any that
could produce this effect.

Thus, on physical grounds, the problem rate should be roughly independent
of separation for $\Delta\theta>30''$.  The mean rate at these separations
is $27/380=7\%$.  The number expected outside $\Delta\theta>60''$ is then
about 15.  Hence, the 10 actually counted is low by only $1.2\,\sigma$.
The analysis of binaries therefore indicates that for separations too 
large to be affected by confusion and fainter than $V>13.5$, the problem 
rate is about $7.1\pm 1.3\%$.  
Inspection of Figures \ref{fig:sep1} and \ref{fig:sep2}
shows that about 2/3 of these problem stars are missing from USNO-B,
and most of the remainder have very significant errors, of order
$100\,\masyr$ or larger.

\section{USNO-B Completeness from Single Stars
\label{sec:sing}}

To find the incompleteness rate among single stars, I search for USNO-B
matches to rNLTT stars whose proper motions are determined by finding
counterparts in USNO-A and 2MASS.  That is, all PPM detections, all
USNO-A-only and all 2MASS-only detections are eliminated.  I also
eliminate all binaries and all stars that are regarded as single by
NLTT but are resolved in TDSC.  These cuts define an rNLTT sample of
20,798 stars.  As summarized in \S~\ref{sec:rnltt},
the proper motions for these stars are accurate to $5.5\,\masyr$,
with very few outliers beyond $30\,\masyr$.  The overall misidentification
rate of the non-PPM stars in rNLTT is about 1\%, but is probably
significantly lower for this subset because it excludes stars that
were identified in only one catalog.  The errors in the 2000-epoch positions
are dominated by 2MASS astrometry errors, and are therefore typically
$<0.\hskip-2pt ''2$.  

I again begin by searching within a $1''$ radius of rNLTT stars, 
which yields 18,420 unique matches and 446 double matches and 1 triple
match.  For the multiple matches, I choose the better match by hand.
I then search among the 1931 non-matches for USNO-B entries lying within a 
$3''$ circle.  For this larger radius, I accept only those stars with proper 
motions within $200\,\masyr$ of rNLTT or (in 11 cases) for which USNO-B has a 
flag  indicating a probable match with NLTT.  This procedure recovers a total
of 130 matches, all unique.

\subsection{Comparison by Magnitude
\label{sec:magcomp}}

Figure \ref{fig:badmag} shows the fraction of problem stars as a function
of $V$ apparent magnitude.  The fraction that are ``bad'' or missing is 
denoted by a solid circle whose error bar is estimated from Poisson (actually
binomial) statistics.  The fraction that are simply missing from USNO-B is
denoted by a cross.  Here ``bad'' means that the amplitude of the
vector difference of the USNO-B and rNLTT proper motion exceeds $30\,\masyr$
and also exceeds twice the USNO-B proper-motion errors.  Recall that
very few rNLTT stars have errors as large as $30\,\masyr$ \citep{faint}.
Among those designated ``bad'', 3/4 have proper-motion differences exceeding
$40\,\masyr$ and 1/2 exceeding $100\,\masyr$.  

	The results presented in Figure \ref{fig:badmag} are broadly 
consistent with those discussed in \S~\ref{sec:bin} but contain more
detail.  The fraction of USNO-B stars that are either unmatched or have
bad proper motions is consistently $\sim 10\%$ over the magnitude range
$14.5\leq V \leq 18.5$.  This is consistent, within statistical 
uncertainties, with the estimate $7.1\pm 1.3\%$ from wide binaries.  Also
in agreement with the binary analysis is the fact that about 2/3 of these
problem stars are unmatched and the remaining 1/3 have large proper-motion
errors.  However, the much better statistics allow one to discern that
the problems with bright stars actually extend beyond the $V=13.5$
boundary inferred from Figure \ref{fig:sep1} to the $V=14$ bin.  Finally,
there is a clear worsening in the final bin, $V=19$, both for missing
and large-error stars.  One might be concerned that the rNLTT errors
themselves grow worse at these faint magnitudes which, if true, would
artificially inflate the number of USNO-B stars with apparently large errors.
However, about 75\% of the ``bad'' USNO-B proper motions have discrepancies
in excess of $100\,\masyr$, and these offsets are unlikely to be due to
rNLTT errors.

	One may be concerned that the results shown in Figure
\ref{fig:badmag} are biased because the rNLTT sample does not include
NLTT stars that lacked counterparts in USNO-A.  If such non-identifications are
correlated between USNO-A and USNO-B, then the resulting incompleteness
estimates would be too low.  In fact, this is a potential concern only
for the last two bins, $V=18$ and $V=19$.  For $13\leq V\leq 17$, the
2MASS-only entries comprise less than 1\% of rNLTT stars, so the
correction is much less than this value.  For the final two bins,
the 2MASS-only entries comprise respectively 3\% and 33\% respectively.
In each case, about 22\% of these 2MASS-only stars are missing, while
about 50\% are either missing or have discrepant proper motions.  Even
if these designations were taken at face value, the fraction of problem
stars in the $V=18$ bin would only rise from 10\% to 11\%, which is barely
significant.  By contrast the fraction of problem $V=19$ stars would rise
from 18\% to 30\%.  However, the characterization of the 2MASS-only stars
must be interpreted very cautiously.  In contrast to the binary case, we
have no independent confirmation that the 2MASS-only identifications are
real.  And if they are, we have no assurance that the adopted NLTT proper
motions are correct.  While \citet{faint} showed that the total fraction
of false identifications in rNLTT is $\sim 1\%$, these are very likely
concentrated in the 2MASS-only detections.  Since less than 3\% of the
rNLTT sample has 2MASS-only identifications, no strong conclusion can
be drawn from the higher rate of USNO-B ``incompleteness'' of these stars.
In addition, for the case of binaries (for which we do have independent
checks on the reality of the identifications), there is no evidence for
an especially high incompleteness rate for 2MASS-only rNLTT identifications
at $V\sim 19$.  In brief, the incompleteness estimates shown in 
Figure~\ref{fig:badmag} are probably not seriously affected by the exclusion
of 2MASS-only stars from the sample.

\subsection{Comparison by Proper Motion
\label{sec:pmcomp}}

Figure \ref{fig:badpm} show that USNO-B incompleteness gradually rises toward 
higher proper motion.  This is to be expected from the intrinsic
difficulty of identifying high proper-motion stars on plates separated
by one to several decades.  It is somewhat exacerbated by the fact that
USNO-B does not attempt to find counterparts whose plate positions differ
by more than $30''$ at different epochs.  Hence, very high proper-motion
stars can be lost simply because they have moved too far between epochs
to be recovered.  Note that the fraction of bad proper motions actually
shrinks toward high proper motions.  That is, if the star is recovered
at all, its proper-motion measurement is rarely seriously in error.

\subsection{Comparison by Galactic Latitude
\label{sec:latcomp}}

Figure \ref{fig:badlat} shows USNO-B incompleteness relative to rNLTT
as a function of (the sine of) Galactic latitude. The sample is restricted
to the magnitude range $13.5<V<18.5$ for which overall completeness
is a maximum (see Fig.~\ref{fig:badmag}).  As would be expected,
incompleteness peaks sharply at the Galactic plane.  

However, Figure \ref{fig:badlat} should be interpreted cautiously.
NLTT (and so rNLTT) is seriously incomplete near the plane at these
faint magnitudes, increasing to $\sim 90\%$ for $15<V<19$ within
$|b|<5.7$ of the plane\footnote{
Fig.\ 14 of \citet{faint} is improperly labeled.  In fact, the solid
(thin) histograms in the upper and middle panels refer respectively to 
main-sequence and subdwarf stars in the mag range $V>15$.
}.
Since it is not known precisely what distinguishes the very few stars 
found by Luyten from the vast majority that he missed, one also does not
know whether these are more or less likely to have been found by USNO-B.
For example, Luyten's completeness was higher for LHS than NLTT.
See for example the comparison of the studies by \citet{monet} and
\citet{nreid} given in the appendix to \citet{parms}.  Since USNO-B
is more incomplete for these higher proper-motion stars 
(see Fig.~\ref{fig:badpm}), this effect would tend to cause USNO-B 
incompleteness to be overestimated.  On the other hand, Luyten may have
preferentially recovered stars that were intrinsically easier to find,
such as brighter stars. If USNO-B also had an easier time recovering
such stars, this would cause USNO-B incompleteness to be underestimated.

We can get some handle on the latter problem by considering the sample
of 141 non-NLTT stars satisfying $|b|<25$, 
$500\,\masyr < \mu < 2000\,\masyr$, $\delta>-2.8$, and $9<R<20$,
found by \cite{lepine}.  In order to compare as directly as possible with
the central two bins of Figure \ref{fig:badlat}, I further restrict this
sample to the 40 (non-binary) stars satisfying $|b|<5.7$ and $14<R<18$.  Of 
these 20 have good USNO-B matches and another 4 have matches with proper-motion
discrepancies larger than $30\,\masyr$.  That is 40\% are missing, and
$50\pm 8\%$ are either missing or bad.  These numbers are somewhat higher
than the two central bins of Figure \ref{fig:badlat}.  However, given that
USNO-B incompleteness is already known to increase at these higher
proper motions (see Fig.~\ref{fig:badpm}), the results are roughly
consistent.

\cite{lepine} suggest that their survey may be 99\% complete 
for $9<R<19$ based on its efficiency recovering NLTT and LHS stars.  
If this were true, then the above analysis would imply that USNO-B
is about 60\% complete even very close to the plane.  However,
\citet{lepine} also allow that NLTT has preferentially found the
easiest stars, so that no strong conclusion can be derived from the
fact that their survey also recovered these.  Indeed, this is very
likely the case, as one may verify by comparing figure 13 of \citet{faint}
with figure 3 of \citet{lepine}.  The former shows that the NLTT ``luminosity
function'' for the whole sky begins falling only for $R>17.5$, while the
latter shows that the luminosity function of \citet{lepine} stars begins
falling for $R>14$.  Thus, at the present time, it is simply not
possible to draw strong conclusions about the completeness of USNO-B
for faint stars close to the Galactic plane.  


\section{Proper-Motion Errors
\label{sec:errors}}

Figure \ref{fig:pmdif2} shows the magnitudes of the vector differences
of three different proper-motion measurements: the abscissa is the
difference between rNLTT and NLTT while the ordinate is the difference
between rNLTT and USNO-B.  The figure has three prominent features: a dense 
horizontal plume along the $x$-axis extending out to $\sim 115\,\masyr$,
a much less dense vertical plume centered at $\sim 30\,\masyr$, and
a still less dense vertical plume centered at $\sim 90\,\masyr$.  The
first vertical plume extends down to the horizontal plume, but the
second appears to cut off below $\sim 150\,\masyr$.

The horizontal plume occurs because there is a tail of NLTT proper-motion
errors extending out to $100\,\masyr$ and beyond.  USNO-B confirms that the
rNLTT measurements are correct, which is what gives this plume its
horizontal character.  \citet{faint} generally searched for 2MASS counterparts
to USNO-A candidates only in a $5''$ radius around the position predicted
using the NLTT proper motion.  Since the typical epoch difference between
these surveys is $\sim 43\,$yr, the maximum proper-motion difference is
$5''/43\,{\rm yr}\sim 115\,\masyr$, which accounts for the cutoff.  

The first vertical plume is due to large USNO-B errors.  The fact that
the plume is centered at $\sim 30\,\masyr$, which is the size of the
(two-dimensional) NLTT proper-motion errors \citep{faint}, means that
NLTT confirms the rNLTT measurements.  About 37\% of this plume is made up
of bright $(V<14.5)$ stars, although these comprise only 22\% of the whole
sample.  Thus, large errors are overrepresented at bright magnitudes but
extend to faint stars as well.  This was already apparent from 
Figure~\ref{fig:badmag}.

The second plume is due to false identifications in rNLTT, i.e. pairs of
unrelated USNO-A and 2MASS stars that happened to have approximately
the vector separation predicted by the NLTT proper motion.  These tend
to be found near the edge of the $5''$ search circle (which contains the
most area).  About half of these points have $V>18.5$ despite the fact that
these faint stars comprise only 6\% of the sample.  Such false identifications
are most likely when the NLTT star is absent from USNO-A, and this occurs
most frequently for faint stars.  The plume appears to begin at $150\,\masyr$
because the 2MASS star does not have a high proper-motion.  Hence, the
ordinate is roughly equal to the rNLTT proper motion whose scale is
set by the proper-motion limit of NLTT, $180\,\masyr$.

I determine the USNO-B errors as a function of $V$ magnitude by
calculating the rms difference between the rNLTT and USNO-B measured
proper motions and subtracting (in quadrature) the rNLTT errors,
$\sigma_\mu({\rm rNLTT})= 5.5\,\masyr$.  The resulting external estimates
per component of the USNO-B errors are shown as open circles in
Figure~\ref{fig:pmerr}.  These estimates are derived by excluding 
$3\,\sigma$ outliers, which constitute roughly 5\% of the sample
for $13\leq V\leq 18$ and 12\% for the remaining three bins.
These external errors may be compared with the rms internal
errors shown as crosses, which are computed from the errors tabulated
for the same stars in USNO-B.  The two sets of estimates are in good overall 
agreement, showing that both USNO-B and rNLTT errors are well-estimated
on average.

How good are the individual USNO-B error bars?  Figure~\ref{fig:intvsext}
shows external error estimates based on comparison with rNLTT as a 
function of internally reported USNO-B errors.  Again $3\,\sigma$
outliers have been excluded and the $5.5\,\masyr$ errors of rNLTT have
been subtracted in quadrature.  The internal error estimates are
accurate over the range 4--20 $\masyr$.  There appears to be a
threshold of $\sigma_\mu=4\,\masyr$.  USNO-B stars with listed
error estimates below this level should be reset at the threshold.  
Also, stars with
reported errors larger than $20\,\masyr$ may have even larger errors.

\section{Discussion
\label{sec:discuss}}

As discussed in the introduction, USNO-B cannot be consulted directly as a
high proper-motion catalog because of the large number of spurious entries.
Here I briefly outline several approaches to using USNO-B to construct
such a catalog.

\begin{figure}
\plotone{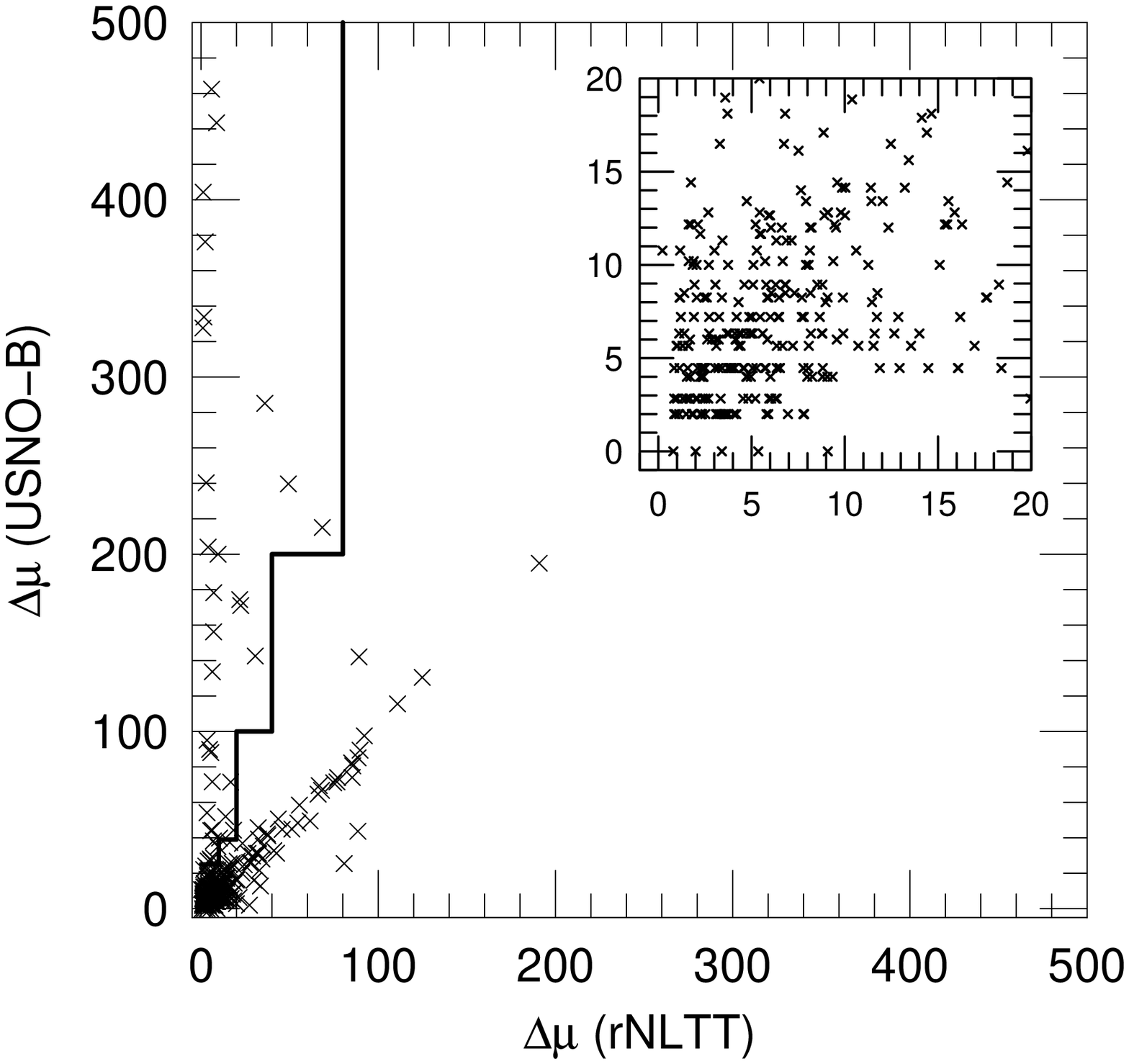}
\caption{\label{fig:pmdif}
Magnitudes of the difference in the vector proper motions of NLTT ``binaries''
as measured by rNLTT \citep{bright,faint} and USNO-B.  The dense knot at
the lower left contains common proper motion (CPM) binaries as determined by
both USNO-B and rNLTT.  The diagonal plume are non-CPM pairs as determined
by both surveys.  The vertical plume are CPM or near-CPM binaries for which
USNO-B shows significantly discrepant proper motions.  These pairs (defined
by the bold boundary) are taken to have at least one 
erroneous USNO-B measurement.
}\end{figure}

\begin{figure}
\plotone{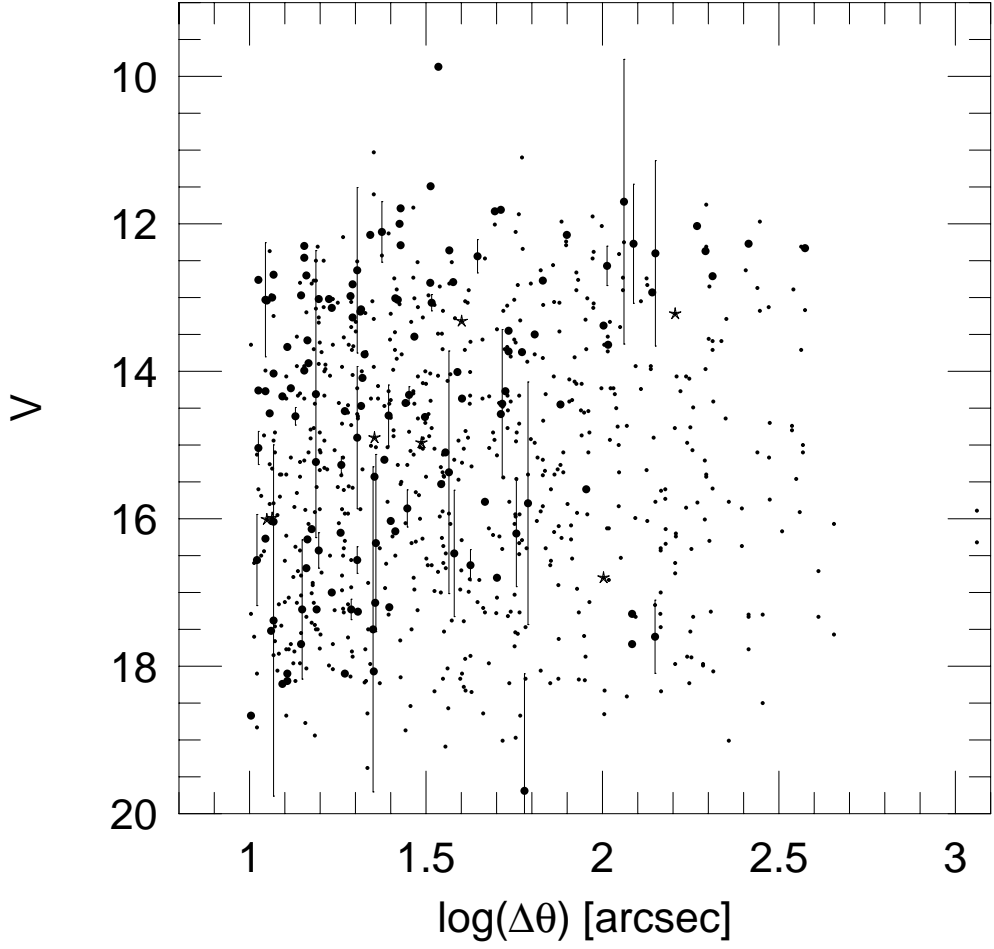}
\caption{\label{fig:sep1}
Recovery of NLTT binary members in USNO-B as functions of
apparent $V$ magnitude and angular separation $\Delta\theta>10''$.  
Small dots represent recovered stars whose proper motion agrees well
with rNLTT (as defined by the bold boundary in Fig.~\ref{fig:pmdif}).  
Filled circles 
are components that are either not found or whose proper motions
do not agree well with rNLTT.  In the latter case, the error
bar indicates the size of the discrepancy in units of
$200\,\masyr$. Star symbols indicate stars recovered by USNO-B whose
companions were not recovered and whose proper motions are
discrepant with rNLTT.  Binaries are included in this plot only if
each component has a proper-motion measurement in rNLTT.  Individual
components whose measurement came from PPM catalogs rather than
USNO-A/2MASS are excluded, since USNO-B did not measure them either.
}\end{figure}

\begin{figure}
\plotone{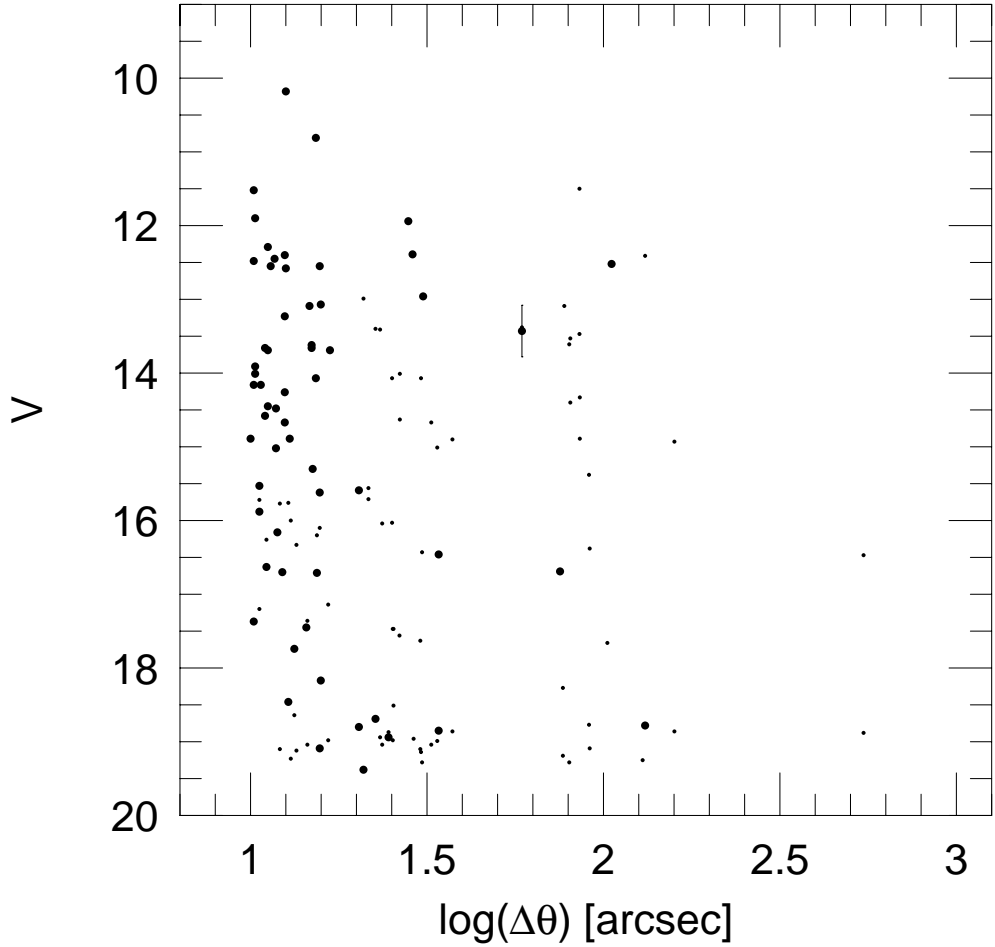}
\caption{\label{fig:sep2}
Similar to Fig.~\ref{fig:sep1} except for binaries that have at least
one component detected by rNLTT in 2MASS only.  For these, proper-motion
information can be obtained by comparing 2MASS-epoch separations with 
those given in the NLTT notes.  Close separations and faint magnitudes
are overrepresented here relative to Fig.~\ref{fig:sep1} because these
stars are preferentially missing from USNO-A.
}\end{figure}

\begin{figure}
\plotone{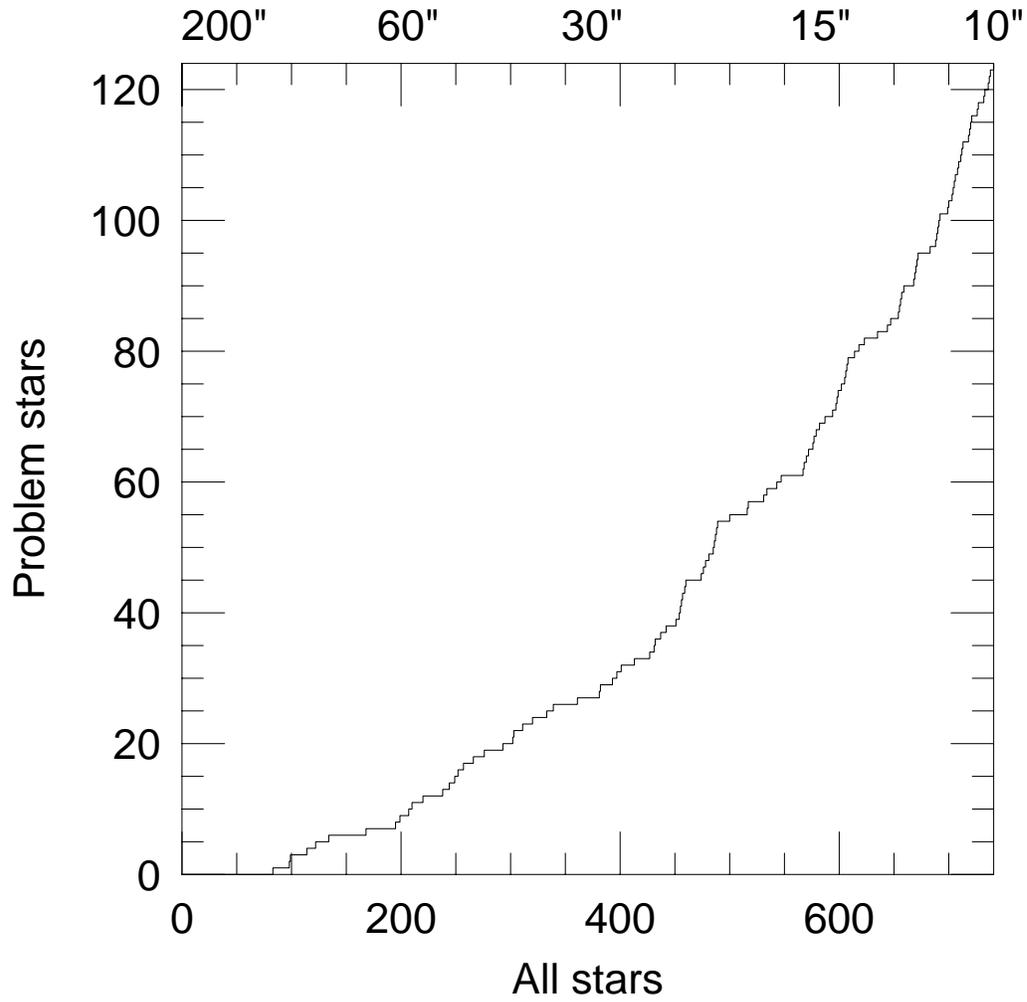}
\caption{\label{fig:cumsep}
Cumulative distribution of problem stars (those missing from USNO-B
or with discrepant proper motions) ranked by binary separation.
There is a clear break at $30''$ due to the onset of confusion caused
by the presence of a companion within this radius.  Apparent breaks
in the curve at wider separations are consistent with statistical
fluctuations.
}\end{figure}

\begin{figure}
\plotone{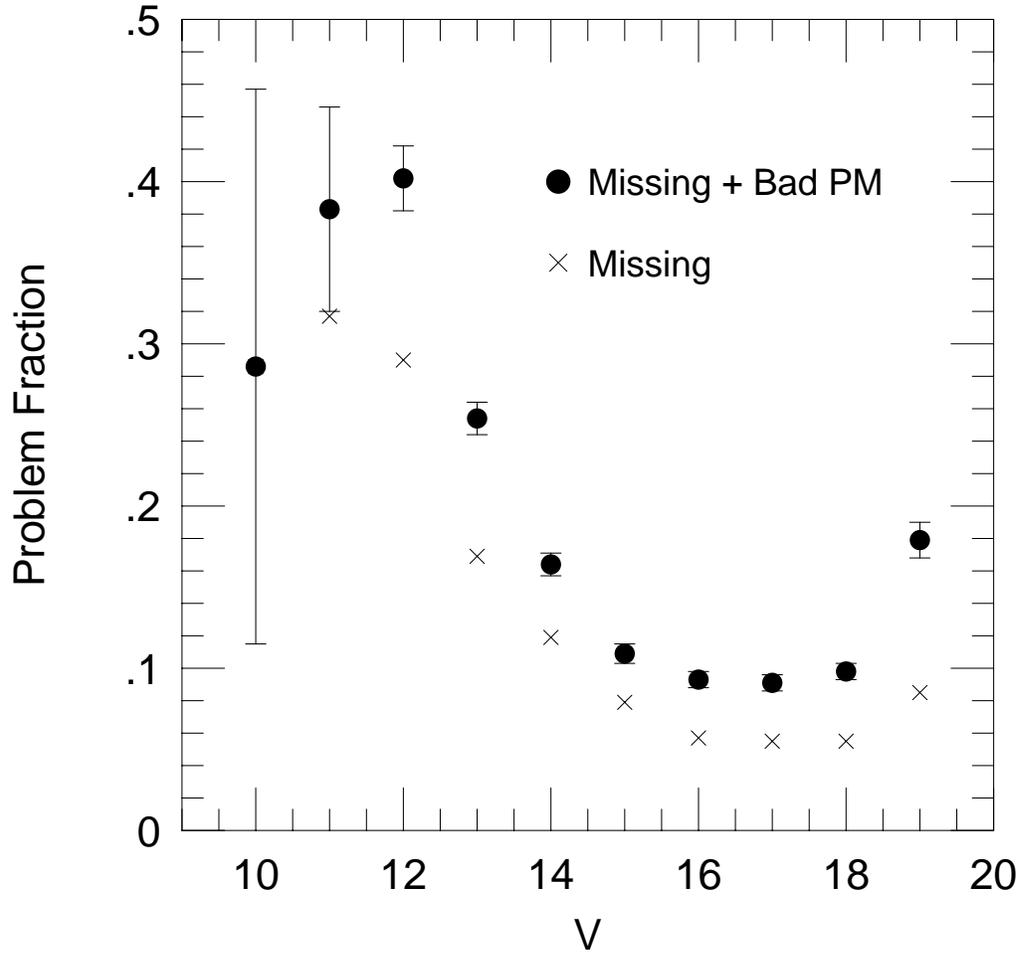}
\caption{\label{fig:badmag}
Incompleteness of USNO-B as a function of $V$ mag for a sample
drawn from rNLTT non-binary stars with identifications in
both USNO-A and 2MASS.  The crosses indicate the fraction that
is strictly missing from USNO-B.  The solid circles add in the
stars whose discrepancy with rNLTT exceeds $30\,\masyr$.
For the majority of these, the discrepancy exceeds $100\,\masyr$.
Error bars are based on counting statistics.
}\end{figure}

\begin{figure}
\plotone{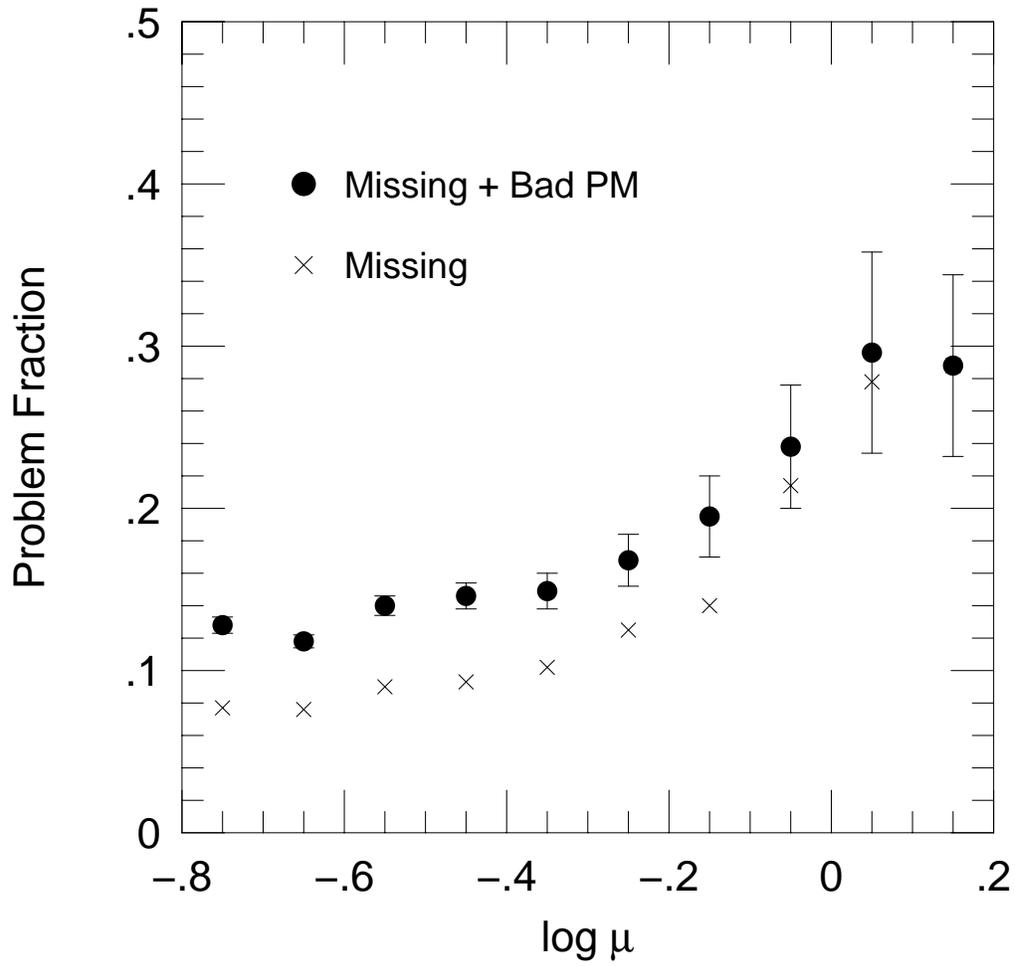}
\caption{\label{fig:badpm}
Incompleteness of USNO-B as a function of proper motion $\mu$.
Symbols are the same as in Fig.~\ref{fig:badmag}
The last bin actually includes all stars $\mu>1259\,\masyr$.
}\end{figure}

\begin{figure}
\plotone{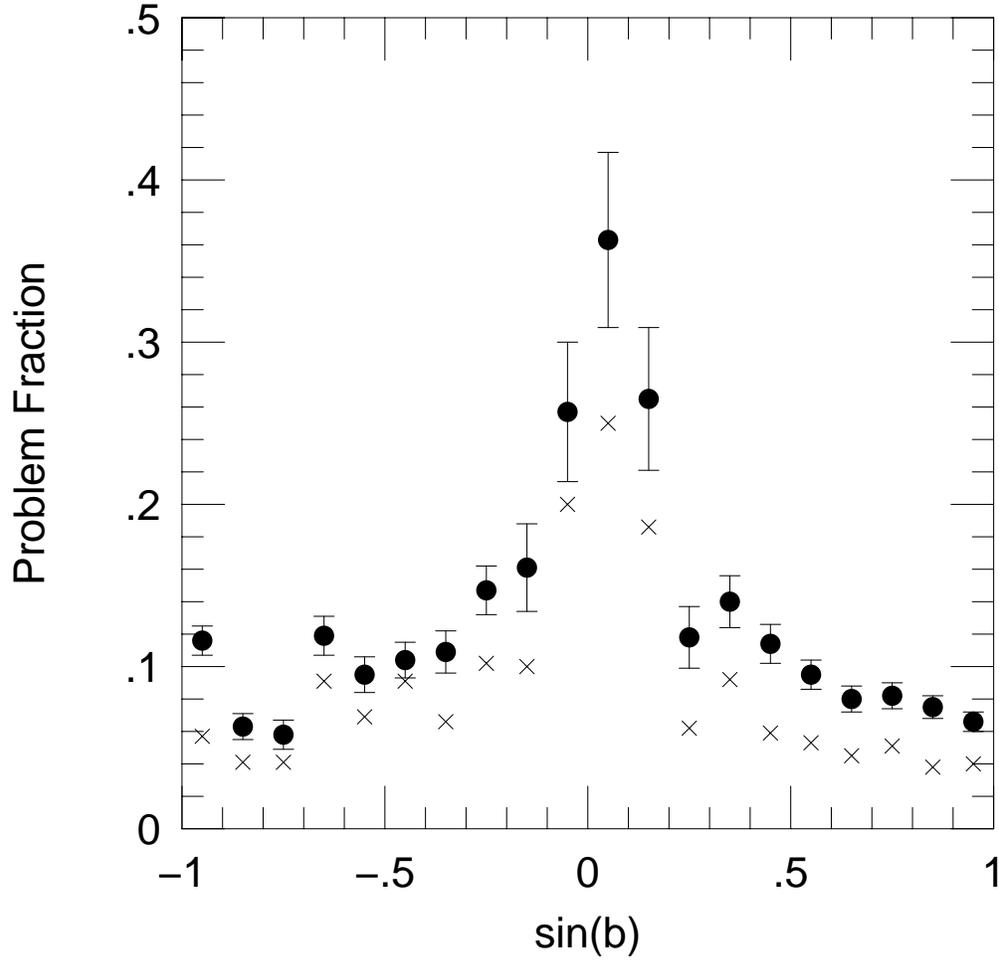}
\caption{\label{fig:badlat}
Incompleteness of USNO-B as a function of Galactic latitude $b$.
Symbols are the same as in Fig.~\ref{fig:badmag}.  Because NLTT
(and so rNLTT) is very incomplete in the plane for $V\ga 14$,
this completeness test applies essentially to brighter stars.
}\end{figure}

\begin{figure}
\plotone{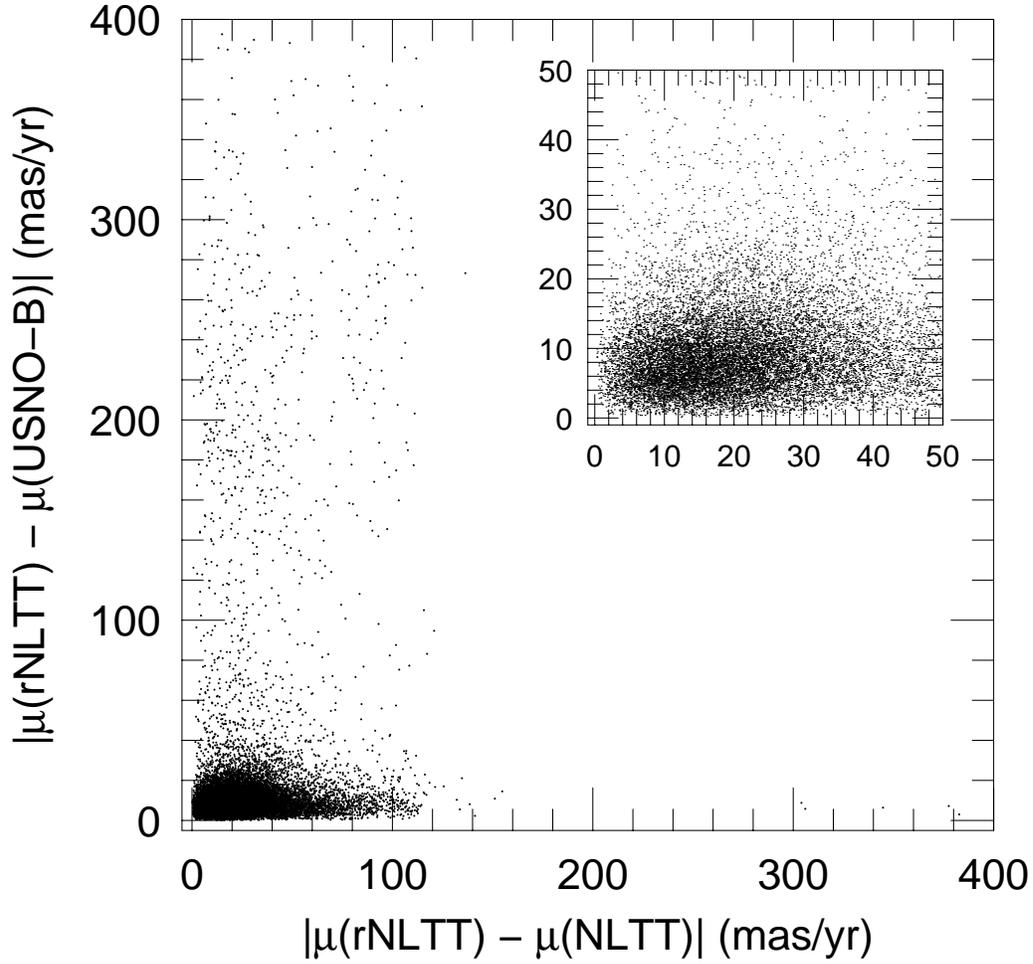}
\caption{\label{fig:pmdif2}
Magnitude of the vector difference of proper motions as given
in 3 catalogs: rNLTT, NLTT, and USNO-B.  The abscissa is the difference
between rNLTT and NLTT while the ordinate is the difference of
rNLTT and USNO-B.  Inset shows an expanded version of the knot at
lower left.  The vertical plume to the left (at $\sim 30\,\masyr$)
is due to errors in USNO-B.  The vertical plume at $\sim 90\,\masyr$
is due to misidentifications in rNLTT.
}\end{figure}

\begin{figure}
\plotone{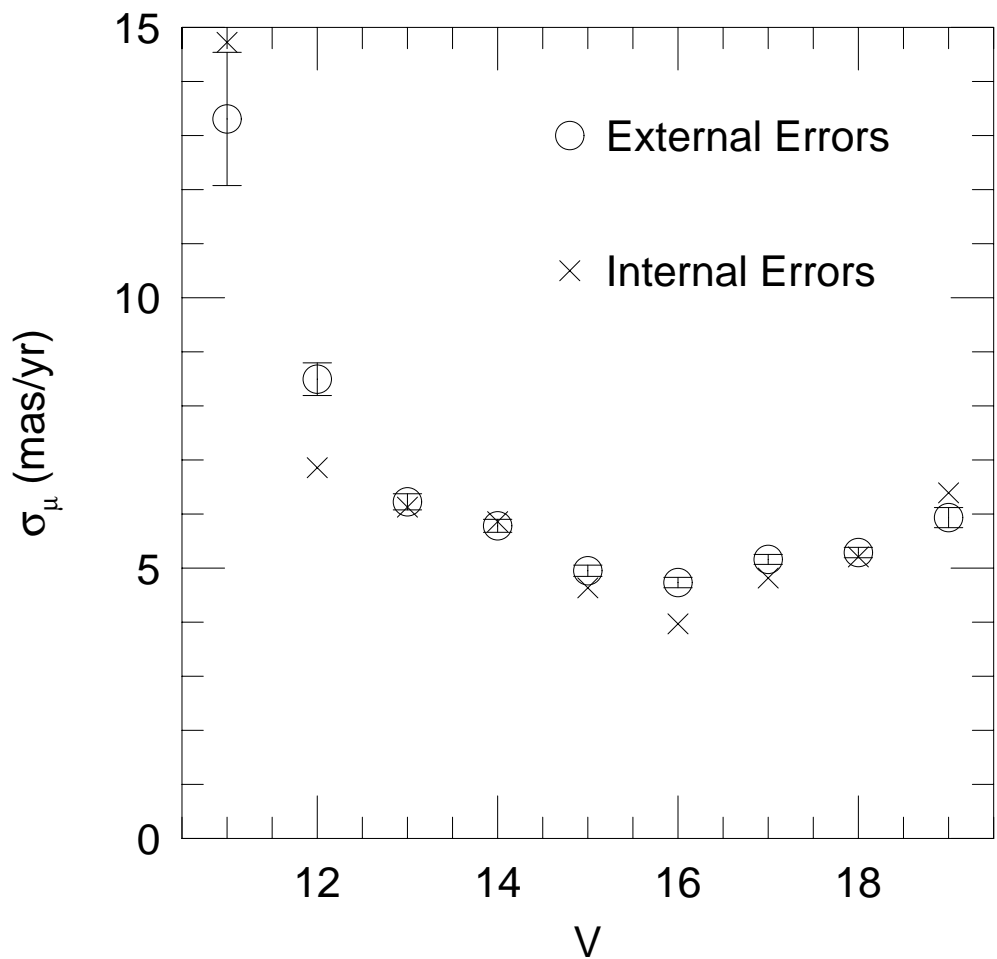}
\caption{\label{fig:pmerr}
Proper-motion errors of USNO-B stars as a function of
$V$ mag.  Open circles show the estimate based on a star-by-star
comparison with rNLTT with $3\,\sigma$ outliers removed and with
the rNLTT error ($5.5\,\masyr$) subtracted in quadrature.  Crosses
show the rms errors of the same stars as reported by USNO-B.  The
error bars are based on counting statistics.  Generally, the agreement
is quite good indicating that, on average, both catalogs have estimated
their errors correctly.  Error bars are based on counting statistics.
}\end{figure}

\begin{figure}
\plotone{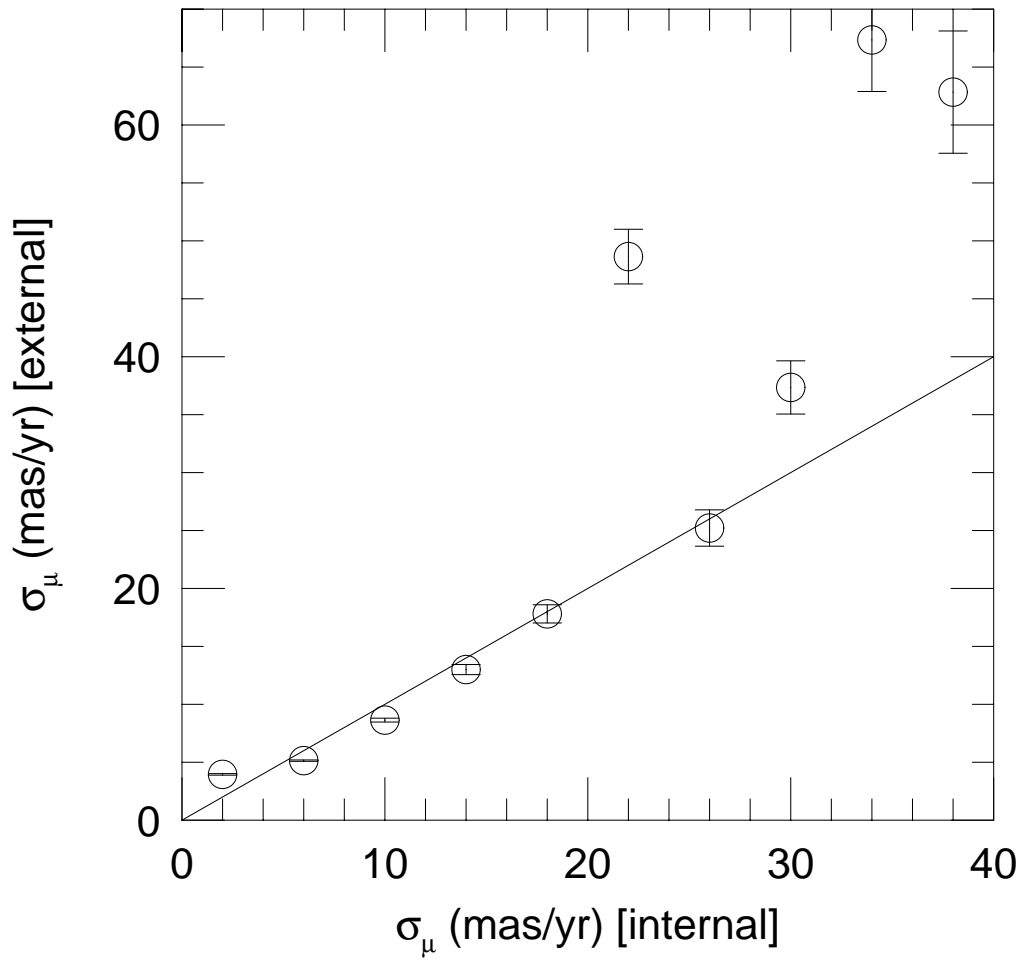}
\caption{\label{fig:intvsext}
Proper-motion errors of USNO-B as a function of internally reported
error estimates.  Line indicates equality.  Agreement is excellent
for $4\,\masyr<\sigma_\mu<20\,\sigma_\mu$.  The leftmost point
probably indicates an error floor of $\sigma_\mu\sim 4\,\masyr$
even when the internally reported errors are below this value.  The
relatively few stars with very large reported proper-motion errors
($\sigma_\mu>20\,\masyr$), may in fact have even larger errors.
}\end{figure}

First, USNO-B can be used to check the identifications of NLTT stars by rNLTT.
While the great majority of the stars in the first vertical plume of
Figure~\ref{fig:pmdif} are USNO-B errors, there must be an extension of
the second plume toward the left, which is then buried in the first plume.
The total number of USNO-B/rNLTT discrepancies is not so great that they
cannot be individually checked.  Indeed if this procedure is extended
to the 2MASS-only and USNO-A-only identifications in rNLTT, it will uncover
a large fraction of the false identifications.

Second, USNO-B can be used to extend rNLTT to the rest of the sky.  For
faint stars ($V\ga 12$) rNLTT presently
covers only the 44\% of the sky defined by the intersection of 
the 2MASS Second Incremental Release and POSS I.
After the full 2MASS release, it will be possible to
apply the approach of \citet{faint} to the rest of the 
POSS I sky ($\delta>-33$),
but this process would be extremely laborious.  It may be far simpler to
simply match USNO-B to 2MASS.  This would eliminate the great majority of
spurious USNO-B entries.  The USNO-B/2MASS matches could then be paired
to NLTT stars using the strategy described in \S~3 of \citet{bright}.
Moreover, this approach could equally well be applied to stars south of
POSS I where the method of \citet{faint} fails because
USNO-A is missing most high proper-motion stars in this region.

The problem of using USNO-B to overcome NLTT's incompleteness is more 
difficult.  Once USNO-B is matched to 2MASS, as described above, one will be 
able to determine the rate of spurious entries by comparing to NLTT.
It may then also be possible to recognize such spurious entries based
on the internal characteristics of the matched entries.  For example,
discrepancies in the photographic magnitudes at various USNO-B epochs,
or inconsistencies between the optical and infrared colors could turn
out to be efficient indicators of spurious entries.  Astrometric 
inconsistencies between USNO-B and 2MASS may also be useful indicators.
Analogs of both these approaches were used by \citet{faint} when matching
2MASS and USNO-A.

More difficult will be the problem of finding high-proper motion stars that
are too faint and/or blue to be present in 2MASS.  For those stars in NLTT,
the above-mentioned approach of \cite{bright} may work even without 2MASS
counterparts.  High proper motion stars that were missed by NLTT and are
absent from 2MASS as well will be most difficult.
Perhaps the most promising prospect is the semi-automatic procedure of
\citet{lepine}.

\acknowledgments 
I thank Samir Salim for his careful reading of the manuscript.
I thank D.\ Monet and the USNO-B team for providing me with a copy of the
USNO-B1.0 catalog.
This work was supported by grant AST 02-01266 from the NSF and JPL contract 
1226901.

\clearpage

\end{document}